\documentclass[USenglish]{article}
\usepackage{natbib}
\usepackage{adjustbox}
\usepackage[small]{dgruyter}
\usepackage{microtype}
\def\aap{A\&A\,  }
\def\aaps{A\&AS  }
\def\apj{ApJ\,  }
\def\apjs{ApJS  }
\def\cjaa{Chinese J. Astron. Astrophys.  }

\def\mnras{MNRAS\,  }

\def\pasp{PASP  }

\def\rmxaa{Revista Mexicana de Astronomia y Astrofisica} 
 
\def\h0units{\mathrm{km\,s^{-1}\,Mpc^{-1}}}
\def\cunits{\mathrm{km\,s^{-1}}}

\newcommand{\om}{\Omega_{\rm M}}
\newcommand{\ok}{\Omega_K}
\newcommand{\ola}{\Omega_{\Lambda}}
\def\sun{\hbox{$\odot$}}
\begin{document}
  \articletype{Editorial}

  \author*[1]{Lorenzo Zaninetti }
  \runningauthor{L. Zaninetti}
  \affil[1]
{ 
Physics Department,
 via P.Giuria 1,\\ I-10125 Turin,Italy
}
\title
{
Filaments of galaxies and Voronoi diagrams
}
\runningtitle{Voronoi Diagrams and galaxies}

\abstract
{
The  intersections between  a spherical shell
and the faces  of Voronoi's  polyhedrons are
numerically evaluated.
The nodes of these intersections are the points that share
the same distances from three nuclei.
The  nodes are assumed to be the places where we will find clusters of
galaxies.
The cosmological environment  is given by the
coupling between  the number of galaxies
as function of the redshift and
the LCDM cosmology.
}

  \startpage{1}
  \aop

\maketitle

\section{Introduction}

The first catalog
of  clusters of galaxies  contained
2712  clusters \cite{Abell1958}.
The updated version  brought the number of rich clusters to
4073, each having at
least 30 members \cite{Abell1989}.
The second  catalog of galaxies contained
29418 galaxies and 9134 clusters
and was organized in six books
\cite{Zwicky1,Zwicky2,Zwicky3,Zwicky4,Zwicky5,Zwicky6}.
The online version of the Zwicky catalog of galaxies
contains 19369  galaxies \cite{Falco1999}.
Another line  of research
organized  the observations
of galaxies in slices; the first being the second CFA2 redshift   Survey
\cite{Huchra1999}.
Other catalogs made by  slices  or which can be organized in slices are:
the 2dfGRS \cite{Norberg2002};
the 6dF Galaxy Survey 6dFGS \cite{Jones2004}); and
the SDSS DR12 with 208,478,448 galaxies \cite{Alam2015}.
All these slices  present filaments of galaxies,
which points toward a cellular structure
for the 3D spatial distribution of galaxies.
A model  for the cellular structure
of the local universe
can be represented in Voronoi Diagrams  \cite{icke1987,Weygaert1989}.
In this model  the galaxies can be inserted on  the
faces of irregular polyhedrons \cite{Zaninetti1991,zaninetti95}.
Recently, the analysis  of  filaments of galaxies
with the Cosmic Web Reconstruction (CWR) has made
enormous progress \cite{Chen2015a,Chen2015b,Chen2016,Chen2017}.

The rest of this  paper is structured as follows. Section \ref{section_cosmology}
introduces the adopted  cosmological framework.
Section \ref{section_voronoi} models the intersection
between a sphere and the network of the Voronoi's faces.
Finally, Section \ref{section_astro} reports on the theoretical
filaments and clusters  of galaxies at low and high values of redshift.

\section{Adopted Cosmology}
\label{section_cosmology}
This section introduces two catalogs of galaxies
(the $\Lambda$CDM and the pseudo-Euclidean cosmology),
the statistics of the cosmic voids
and  the standard luminosity function (LF)  for galaxies.

\subsection{The adopted catalogs}

The 2MASS Redshift Survey (2MRS)
has
44599  galaxies
between  $0<z<0.17$
and  covers $91\%$ of the sky,
see \cite{Huchra2012}.
The  redMaPPer  catalog
has 25000 clusters of galaxies between  $0.08<z<0.55$
and covers  $24.2\%$ of the sky,
see  \cite{Rykoff2014}.

\subsection{Luminosity distance}
We review the existing knowledge on
the luminosity distance, $D_L(z;H_0,c,\om,\ola)$,
which in  the $\Lambda$CDM cosmology can
be expressed in terms of a Pad\'e approximant.
We should provide:
the Hubble constant, $H_0$, as expressed in  $\h0units$;
the velocity of light, $c$, as expressed in $\cunits$; and
the three numbers $\om$, $\ok$, and $\ola$,
(see \cite{Zaninetti2016a} for more details).
A numerical analysis
of the distance modulus for the Union 2.1 compilation, see
\cite{Suzuki2012},
gives $H_0 = 69.81 \h0units$, $\om=0.239$  and  $\ola=0.651$.
We now apply     the minimax rational approximation,
which is characterized by  two parameters $p$ and $q$,
and we  find    a simplified expression
for the luminosity distance, $D_{L,2,1}$,
when $p=2$ and $q=1$
\begin{eqnarray}
D_{L,2,1}=
\frac
{- 151.187045+ 4991.403961\,z+ 1643.868844\,{z}^{2}  }
{ 0.9262689263+ 0.03596637742\,z }
\\
\quad  for \quad 0 <z<4
\quad .
\nonumber
\label{dlz}
\end {eqnarray}
This  equation  can be inverted to give
a "{\it new}" expression for 
the redshift  as a function
of the luminosity distance in Mpc
\begin{eqnarray}
z = 1.09395\times 10^{-5} \,D_{{L,2,1}}- 1.51818
+{ 6.08321\times 10^{-15}} \times
\nonumber \\
\sqrt { 3.23395\times10^{18} \,{D_{{L,2,1}}}^{2}+{ 1.4329
\times 10^{25}}\,D_{{L,2,1}}+{ 6.47706\times 10^{28}}}
\label{zdl}
\\
\quad for\quad  0 < D_{{L,2,1}}    < 43094 \quad .
\nonumber
\end{eqnarray}
The most simple  model for  the  distance, $d$,
in the local universe
is that of  the pseudo-Euclidean cosmology:
\begin{equation}
d(z;c,H_0)=
\frac {z c}{H_{{0}}}
\quad ,
\label{distancepseudo}
\end{equation}
where we used $H_0=67.93 \h0units $, see \cite{Zaninetti2016c}.

\begin{figure}
\begin{center}
\includegraphics[width=10cm]{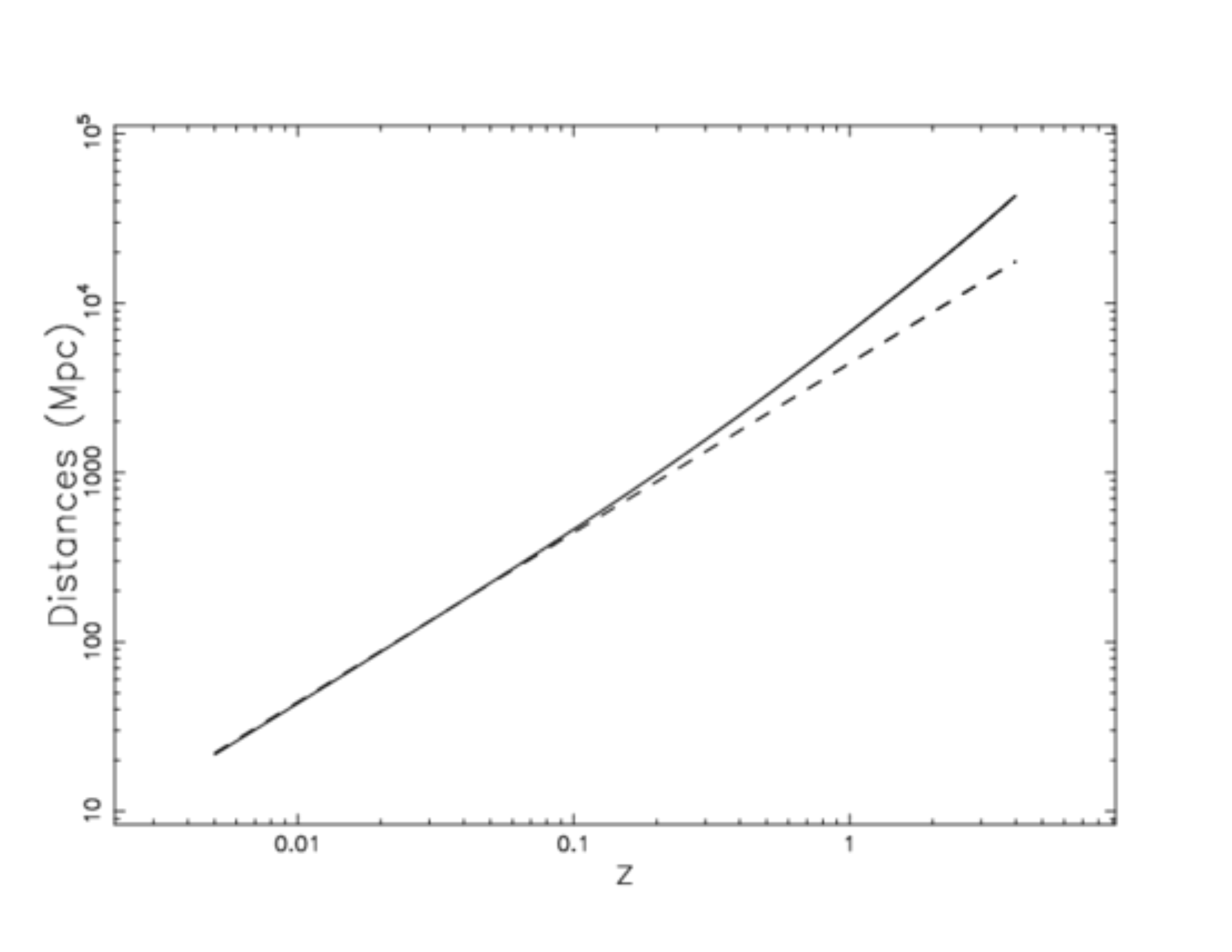}
\end{center}
\caption
{
The  distances that are adopted are:
the luminosity distance,        $D_L$, in $\Lambda$CDM (full line);
and
the pseudo-Euclidean cosmology distance (dashed  line).
}
\label{distances}
\end{figure}

The differences between the two distances  are the
luminosity distance and the
and the
pseudo-Euclidean  distance, which can be outlined
in terms of a percentage difference:
 $\Delta$.
For example, for $ d$,
\begin{equation}
\Delta = \frac{\big | D_{\rm L}(z) - d (z) \big |}
{D_{\rm L}(z)} \times 100
\quad .
\end{equation}
Figure \ref{distances} reports the two distances.
For $z\leq 0.2$, the  percentage difference is
lower than $10\%$.

Therefore the boundary between low and high $z$ can be fixed
at $z=0.2$.

\subsection{Cosmic voids}

A {\it first} catalog of cosmic voids
can be found
in \cite{Vogeley2012},
where the effective radius of the voids, $R_{eff}$,
has been derived
\begin{equation}
R_{eff} = 18.23 h^{-1}\, Mpc  \quad Pam\, et \, al. \,2012
\quad .
\end{equation}
The {\it second} catalog is that of  radii
up to redshift
$0.12\,h^{-1} Mpc$   in (SDSS-DR7), see \cite{Varela2012},
\begin{equation}
R_{eff} = 11.85 h^{-1}\, Mpc \quad  Varela\, et \, al. \,2012
\quad.
\end{equation}
The {\it third}  catalog  is that  of the
Baryon Oscillation Spectroscopic Survey,
see \cite{Mao2017},
\begin{equation}
R_{eff} = 57.53 h^{-1}\, Mpc \quad  Mao\, et \, al. \,2017
\quad.
\end{equation}
In the following, we will calibrate our code
on the average value  of the three previous values:
$\overline{R_{eff}}=29.2\,h^{-1}\,Mpc$.

\subsection{The LF for galaxies}

We now review the actual knowledge on  
the  Schechter function,
see  \cite{schechter} which  
provides a useful standard   for the
LF  of galaxies
\begin{equation}
\Phi (L) dL  = (\frac {\Phi^*}{L^*}) (\frac {L}{L^*})^{\alpha}
\exp \bigl ( {-  \frac {L}{L^*}} \bigr ) dL \quad  ,
\label{equation_schechter}
\end {equation}
where $\alpha$ sets the slope for low values
of luminosity, $L$,  $L^*$ is the
characteristic luminosity and $\Phi^*$ is the normalisation.
The equivalent distribution in absolute magnitude is
\begin{equation}
\Phi (M)dM=0.921 \Phi^* 10^{0.4(\alpha +1 ) (M^*-M)}
\exp \bigl ({- 10^{0.4(M^*-M)}} \bigr)  dM \, ,
\label{lfstandard}
\end {equation}
where $M^*$ is the characteristic magnitude as derived from the
data.
The scaling with  $h$ is  $M^* - 5\log_{10}h$ and
$\Phi^* ~h^3~[Mpc^{-3}]$.
According to formula~(1.104) in
 \cite{pad}
or formula~(1.117)
in
\cite{Padmanabhan_III_2002}
,
the joint distribution in redshift, {\it z},
and  flux, {\it f},
for galaxies in the pseudo-Euclidean cosmology,
 is
\begin{equation}
\frac{dN}{d\Omega dz df} =
4 \pi  \bigl ( \frac {c}{H_0} \bigr )^5    z^4 \Phi (\frac{z^2}{z_{crit}^2})
\label{nfunctionzschechter}
\quad ,
\end {equation}
where $d\Omega$, $dz$ and  $ df $ represent
the differential of
the solid angle,
the redshift and the flux, respectively,
and     $\Phi$ is the Schechter LF.
The critical value of $z$,   $z_{crit}$, is
\begin{equation}
 z_{crit}^2 = \frac {H_0^2  L^* } {4 \pi f c^2}
\quad .
\end{equation}
The number of galaxies in $z$  and $f$ as given by
formula~(\ref{nfunctionzschechter})
has a maximum  at  $z=z_{pos-max}$,
where
\begin{equation}
 z_{pos-max} = z_{crit}  \sqrt {\alpha +2 }
\quad ,
\end{equation}
which  can be re-expressed   as
\begin{equation}
 z_{pos-max}(f) =
\frac
{
\sqrt {2+\alpha}\sqrt {{10}^{ 0.4\,{\it M_{\sun}}- 0.4\,{\it M^*}}}{
\it H_0}
}
{
2\,\sqrt {\pi }\sqrt {f}{\it c}
}
\quad  ,
\label{zmax_sch}
\end{equation}
where $M_{\sun}$ is the reference magnitude
of the sun at the considered bandpass.
On   replacing  the flux $f$ with the apparent
magnitude $m$
\begin{equation}
 z_{pos-max}(m) =
\frac
{
1.772\,10^{-5}
\,\sqrt {2+\alpha}\sqrt {{10}^{ 0.4\,M_{{{\it \sun}}}-
 0.4\,{\it M^*}}}H_{{0}}
}
{
\sqrt {\pi }\sqrt {{{\rm e}^{ 0.921\,M_{{{\it \sun}}}-
 0.921\,m}}}{\it c}
}
\quad  .
\label{zmax_schmag}
\end{equation}

\begin{figure}
\begin{center}
\includegraphics[width=10cm]{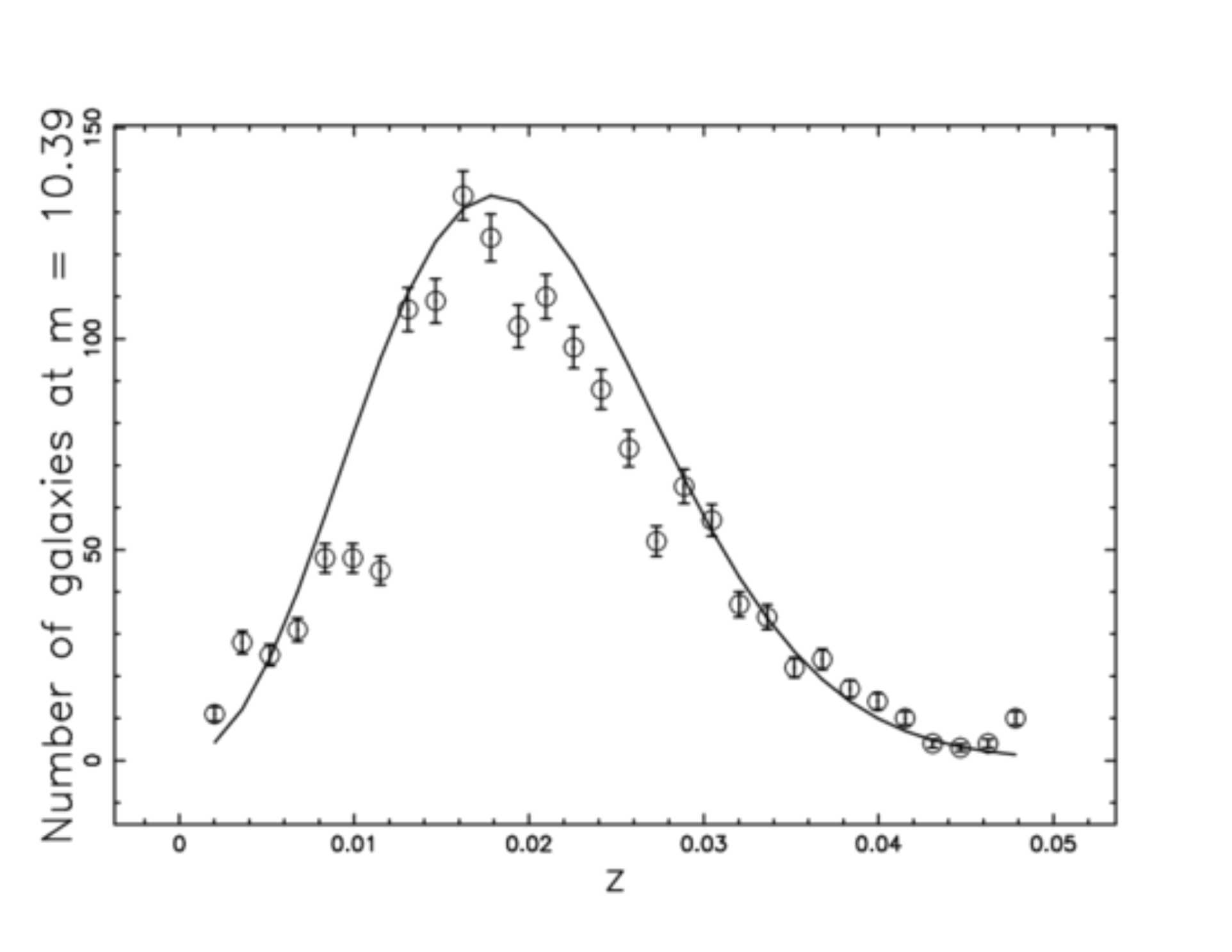}
\end{center}
\caption
{
The galaxies  of the 2MRS with
$ 10.31    \leq  m   \leq 10.47  $  or
$ 1164793  \frac {L_{\sun}}{Mpc^2} \leq
f \leq  1346734 \frac {L_{\sun}}{Mpc^2}$
are  organized in frequencies versus
heliocentric  redshift,
(empty circles); and,
the error bar is given by the square root of the frequency.
The maximum frequency of observed galaxies is
at  $z=0.017$.
The full line is the theoretical curve
generated by
$\frac{dN}{d\Omega dz df}(z)$, see  equation(\ref{nfunctionzschechter}),
and the maximum is at $z=0.0133$.
The parameters are given  in Table \ref{parameterslf}.
}
\label{maximum_2mrs}
\end{figure}

Figure \ref{maximum_2mrs}
reports the number of  observed  galaxies
of the 2MRS  catalog for  a given
apparent magnitude  and for the
theoretical curve.
Table  \ref{parameterslf} reports the parameters that were adopted in this
 model.
 \begin{table}
 \caption[]{
Parameters of the Schechter LF before
scaling with h.
}
 \label{parameterslf}
 \[
 \begin{array}{ll}
 \hline
parameter &  value \\ \noalign{\smallskip}
 \hline
 \noalign{\smallskip}
M^*         & -23.15  \\
\alpha      & -0.78    \\
\Phi^*      & 0.0128 / Mpc^3  \\
 M_{\sun}   & 3.39            \\
 H_0        & 67.93    \,\h0units       \\
\noalign{\smallskip}
 \hline
 \hline
 \end{array}
 \]
 \end {table}

\begin{figure}
\begin{center}
\includegraphics[width=10cm]{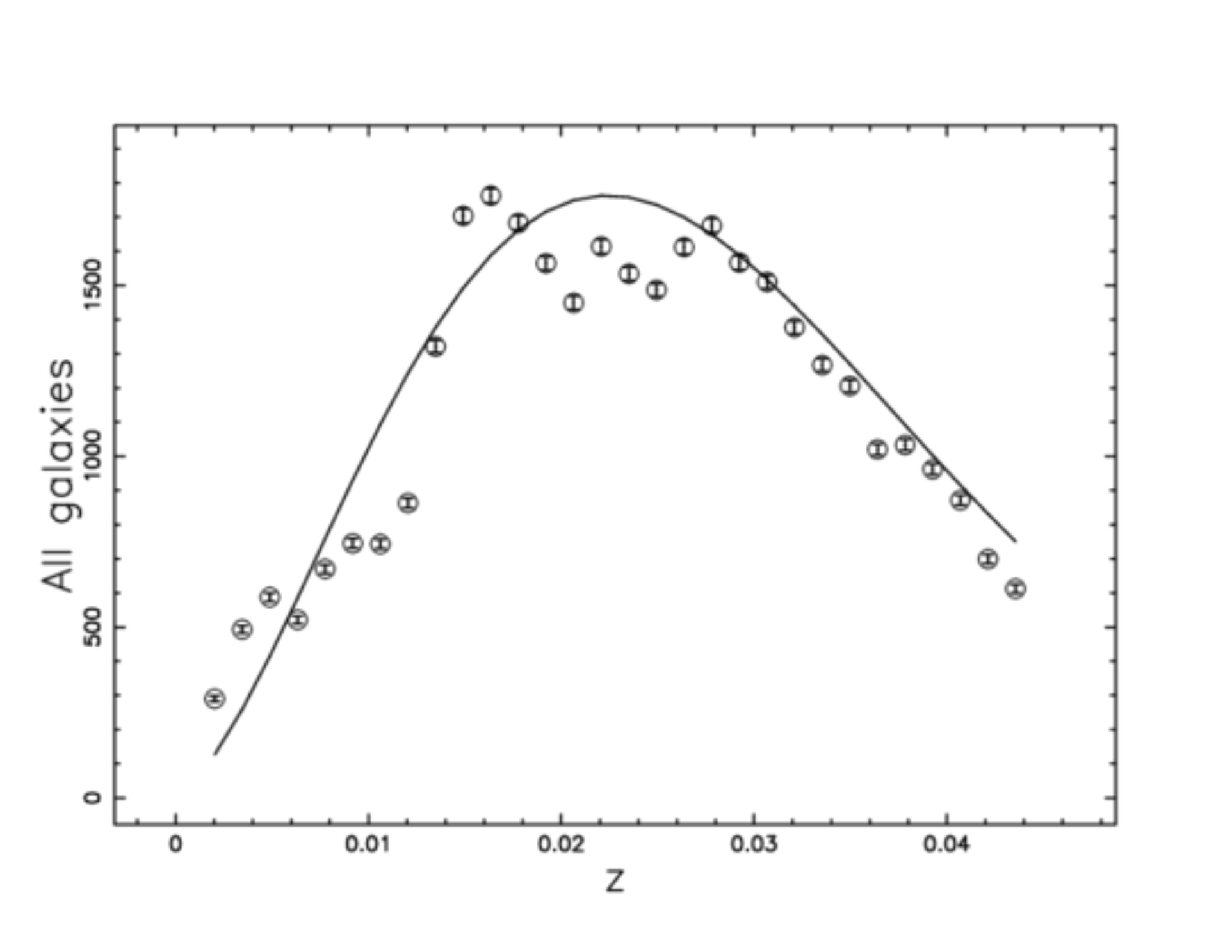}
\end{center}
\caption
{
All the  galaxies  for  2MRS
are  organised in frequencies versus
redshift,
(empty circles);
the error bar is given by the square root of the frequency.
The maximum frequency of the observed galaxies is
at  $z=0.017$.
The full line is the theoretical curve
generated by
the integral   equation (\ref{integraletutte}),
which has  maximum is at $z=0.0237$.
The parameters are given  in Table \ref{parameterslf}.
}
\label{maximum_2mrs_all}
\end{figure}

The number of galaxies, $N(z,f_{min},f_{max})$
comprised between a minimum value of flux,
 $f_{min}$,  and  maximum value of flux $f_{max}$,
can be computed  through  the following integral
\begin{equation}
N (z) = \int_{f_{min}} ^{f_{max}}
4 \pi  \bigl ( \frac {c_l}{H_0} \bigr )^5    z^4 \Phi (\frac{z^2}{z_{crit}^2})
df
\quad .
\label{integrale}
\end {equation}
The indefinite integral exists in terms  of
the Whittaker function ${{\sl
M}_{a,\,b}\left(x\right)}$, see \cite{Abramowitz1965,NIST2010},
but has a complicated expression.
On  inserting
the model's parameters
in this  integral,
we obtain the following "{\it new}" expression
\begin{eqnarray}
N(z)=
9414469\,{z}^{2} \left( {z}^{2} \right) ^{- 0.39}{{\rm e}
^{- 249394.29\,{z}^{2}}}{{\sl M}_{- 0.39,\, 0.11}
\left( 498788.58\,{z}^{2}\right)}
\nonumber \\
- 106747137\,{z}^{2} \left( {z}^{
2} \right) ^{- 0.39}{{\rm e}^{- 493.043\,{z}^{2}}}{{\sl M}
_{- 0.39,\, 0.11}\left( 986.0876\,{z}^{2}\right)}
\quad .
\label{integraletutte}
\end{eqnarray}
The number   of  all of the galaxies for 2MRS as function of the redshift
is visible  in Figure \ref{maximum_2mrs_all}.
More details can be found in \cite{Zaninetti2014b}.

\section{Voronoi Diagrams}
\label{section_voronoi}

We now review the existing knowledge on the Voronoi diagrams. 
The faces of the Voronoi Polyhedra
share the same property; i.e.,
they are equally distant from
two nuclei or seeds.  The intersection
between a plane and the faces produces diagrams that are similar
to the edges   displacement in  2D Voronoi diagrams.
From the point of view of the observations,
it is very useful to study the intersection
between a slice which crosses  the center of the box
and the faces of  irregular polyhedrons where
the galaxies presumably reside.
According to the nomenclature reported
in \cite{Okabe2000},  this cut is classified as
$V_P(2,3)$.
The  parameters that will be used in the following
are the kind  of  nuclei, which can be  Poissonian or not Poissonian,
the number of nuclei,
the $side$ of the box in $Mpc$,
and  the number of $pixels$, for example 1400, that are
used  to build the diagrams,
see \cite{zaninettig}.
The  parameters adopted  in  the  simulation of the
pseudo-Euclidean cosmology  are reported in
Table\ref{parameterspseudo}.
\begin{table}[ht!]
\caption
{
Numerical values
for the parameters of the Voronoi Diagrams.
}
\label{parameterspseudo}
\begin{center}
\begin{adjustbox}{max width=\textwidth}
\begin{tabular}{|l|c|c|c|c|c|c|}
\hline
catalog    & Cosmology & pixels    &type~of~seeds& seeds & box-side [Mpc] & radius [z]  \\
\hline
2MRS       & pseudo-Euclidean & 1400      &Poissonian   & 659  &   441          &  0.05  \\
redMaPPer  & $\Lambda$CDM&  700      &Poissonian   & 3914 &  1060          &  0.115 \\
\hline
\end{tabular}
\end{adjustbox}
\end{center}
\end{table}

\begin{figure}\begin{center}
\includegraphics[width=7cm]{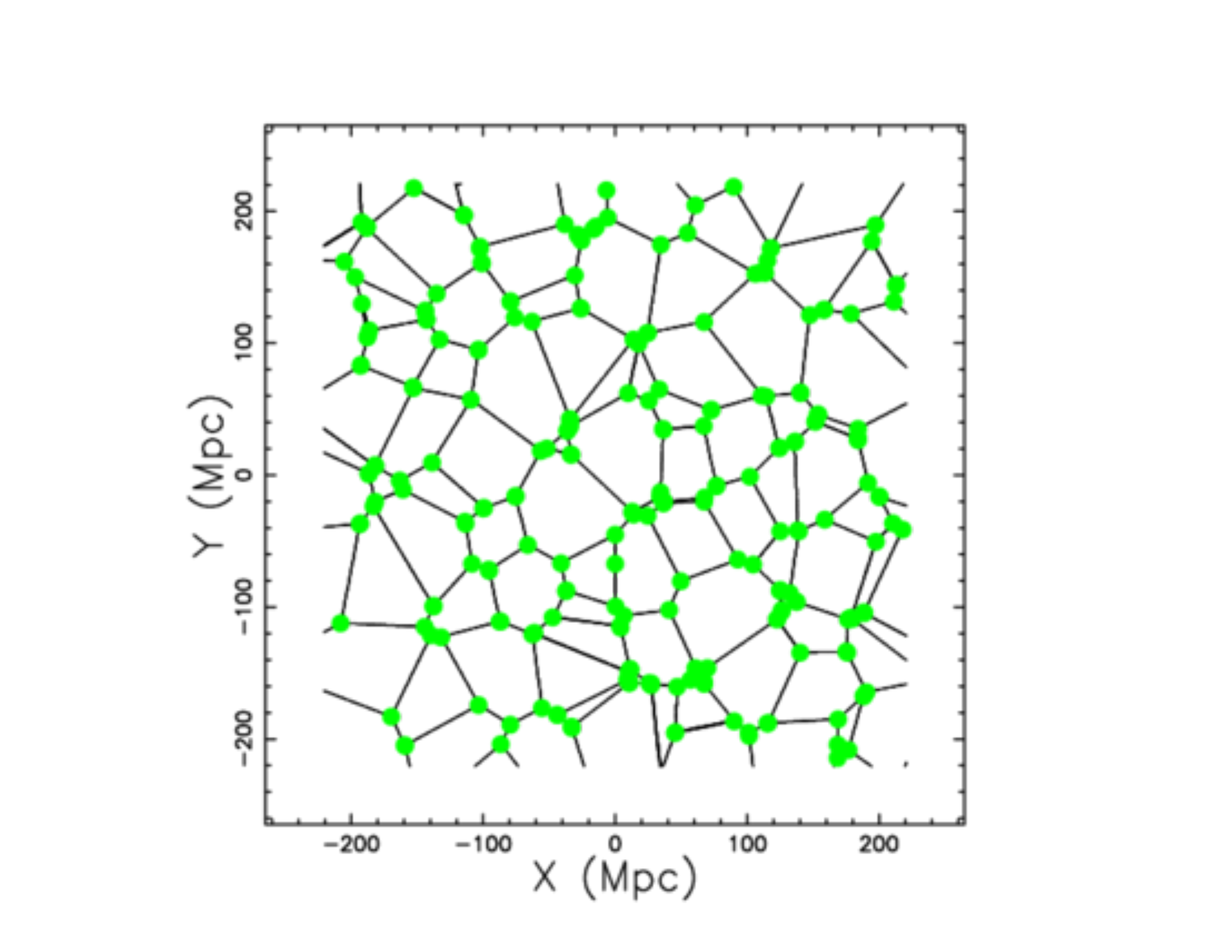}
\end{center}
\caption
{
Portion of the  Poissonian Voronoi diagram $V_P(2,3)$;
cut on the  X-Y plane, black points,
the parameters are given in Table \ref{parameterspseudo}.
The spherical nodes are marked by great green points.
}
\label{nodi_mezzo}%
    \end{figure}

\begin{figure}\begin{center}
\includegraphics[width=7cm]{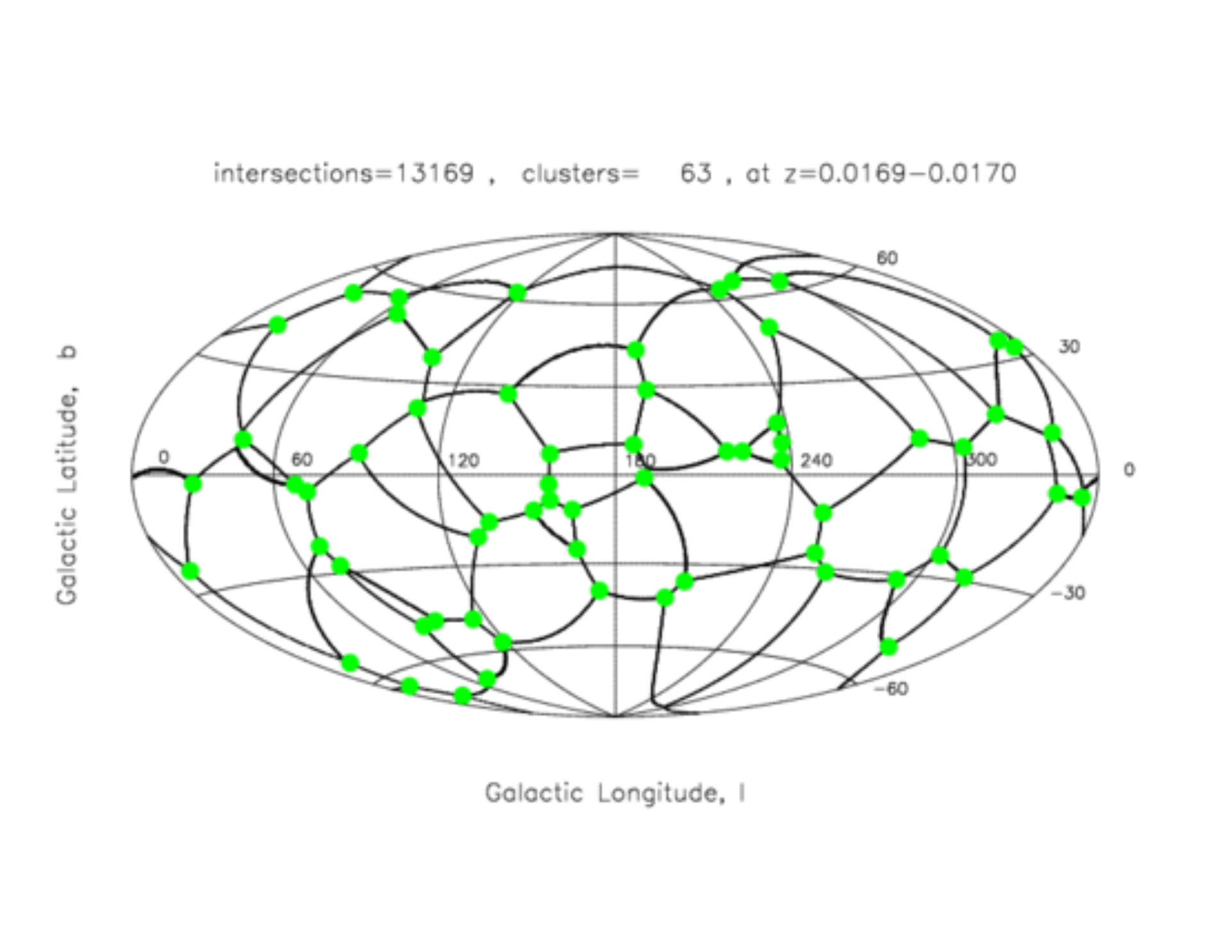}
\end{center}\caption
{
The Voronoi-diagram $V_{P,s}(2,3)$
in the Hammer--Aitoff  projection
at $ 0.0169 < z < 0.0170$
with the same parameters as in Table \ref{parameterspseudo}.
}
          \label{aitof_sphere}%
    \end{figure}

According to the nomenclature reported
in \cite{Okabe2000},  this cut is classified as
$V_P(2,3)$  and  Figure \ref{nodi_mezzo} reports
a typical example.
This Figure also reports  the spherical nodes
that, in the absence of an official definition,
can be defined as the locus of intersection
between the lines of $V_P(2,3)$.
The spherical nodes
are equally distant from
three or four  nuclei.
The cross-sectional area of the VP can also
be visualised through
a spherical cut that is characterised by a constant value
of the distance to the center of the box,
which in this case is expressed in $z$ units.
This intersection
is not present in the Voronoi literature and therefore
can be classified as a "{\it new}" topic.
It may be  called $V_{P,s}(2,3)$,  where the
index $P,s$ stand  for Poissonian and sphere respectively,
see Figure \ref{aitof_sphere}.
More details  can be found in \cite{Zaninetti2006}.

\section{Astrophysical Applications}
\label{section_astro}
This  section reports the theoretical spatial  display
for clusters at low redshift, simulation of 2MRS,
and at high redshift,  simulation of redMaPPer.

\subsection{The 2MRS catalog}

\begin{figure}\begin{center}
\includegraphics[width=7cm]{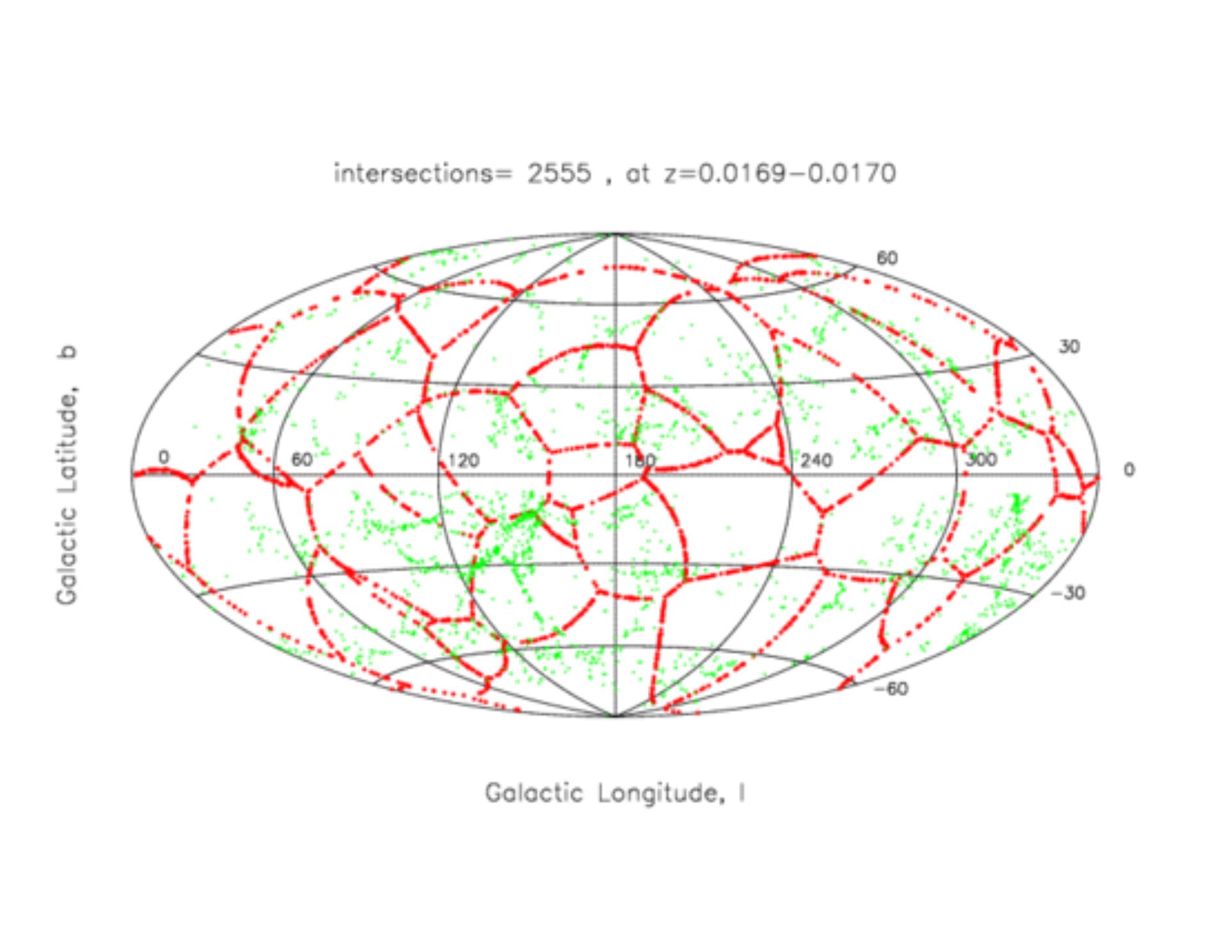}
\end{center}\caption
{
The 2555 galaxies of 2MRS at $ 0.0169 < z < 0.0170$ (green points)
and the same number of $V_{P,s}(2,3)$
in the Hammer--Aitoff  projection (red points).
The parameters are given in Table \ref{parameterspseudo}.
}
          \label{2mrs_aitoff_due}%
    \end{figure}

\begin{figure}\begin{center}
\includegraphics[width=7cm]{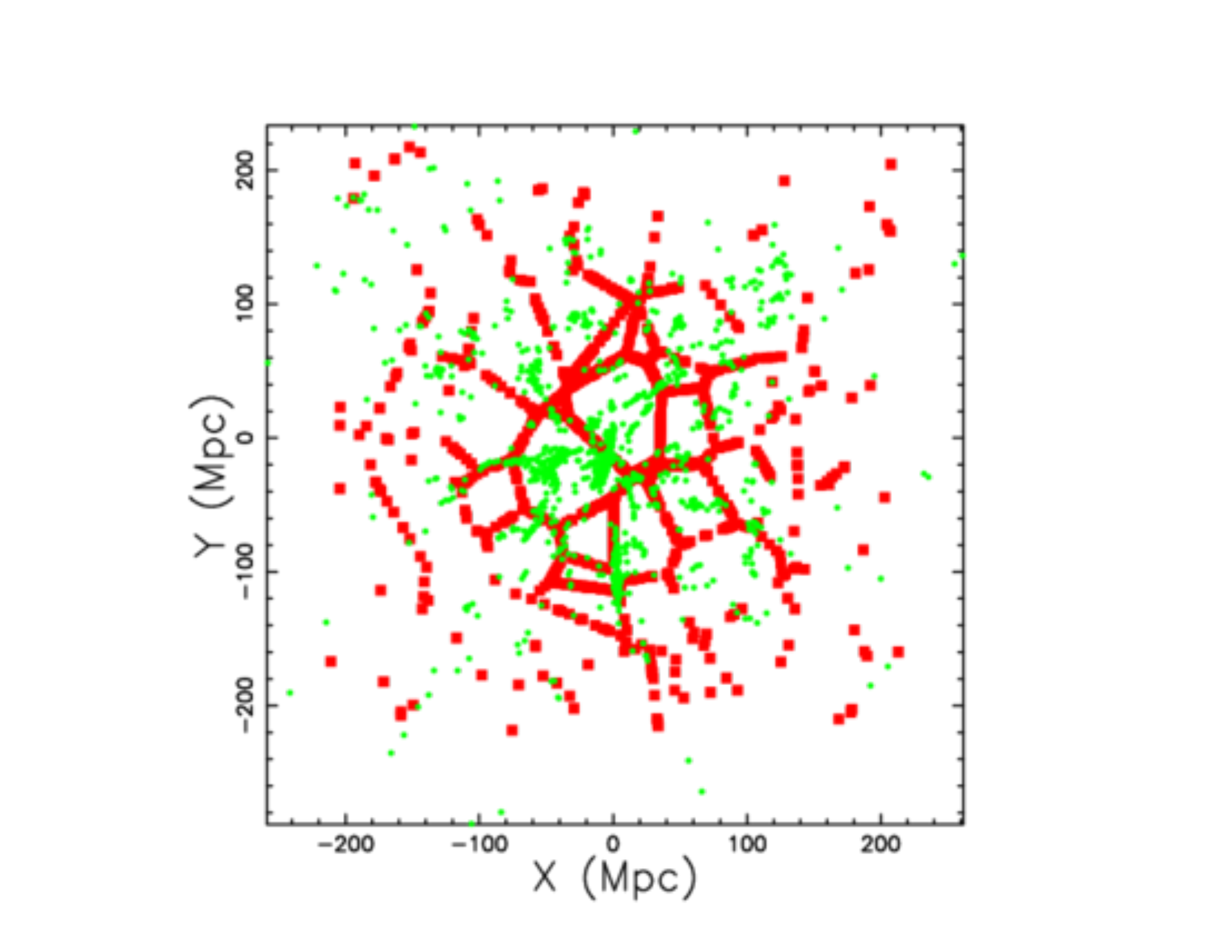}
\end{center}\caption
{
A cut  of the 3D spatial distribution  of 2MRS
in the $Z=0$ plane  when  $\Delta =$ 4 Mpc,
the squared box has a side of  614 $Mpc$:
we have 1384 galaxies (green filled circles).
The same  number of $V_P(2,3)$ (red squares)
with radial  scaling  as the number of real galaxies
with the same parameters as in Table \ref{parameterspseudo}.
}
          \label{2mrs_scaling}%
    \end{figure}

The  photometric maximum
in the number of  galaxies as a function of $z$,
see formula (\ref{integraletutte}),
has  a maximum  as a function of the redshift.
Figure \ref{2mrs_aitoff_due} reports an
Hammer--Aitoff  projection of the  galaxies of 2MRS,
as well the same number of $V_{P,s}(2,3)$.
Figure \ref{2mrs_scaling} reports a  cut
of  a given thickness, $\Delta$,
of 2MRS and  a number of  $V_P(2,3)$ chosen
to scale as the number of galaxies.

\subsection{The redMaPPer  catalog}

\begin{figure}\begin{center}
\includegraphics[width=7cm]{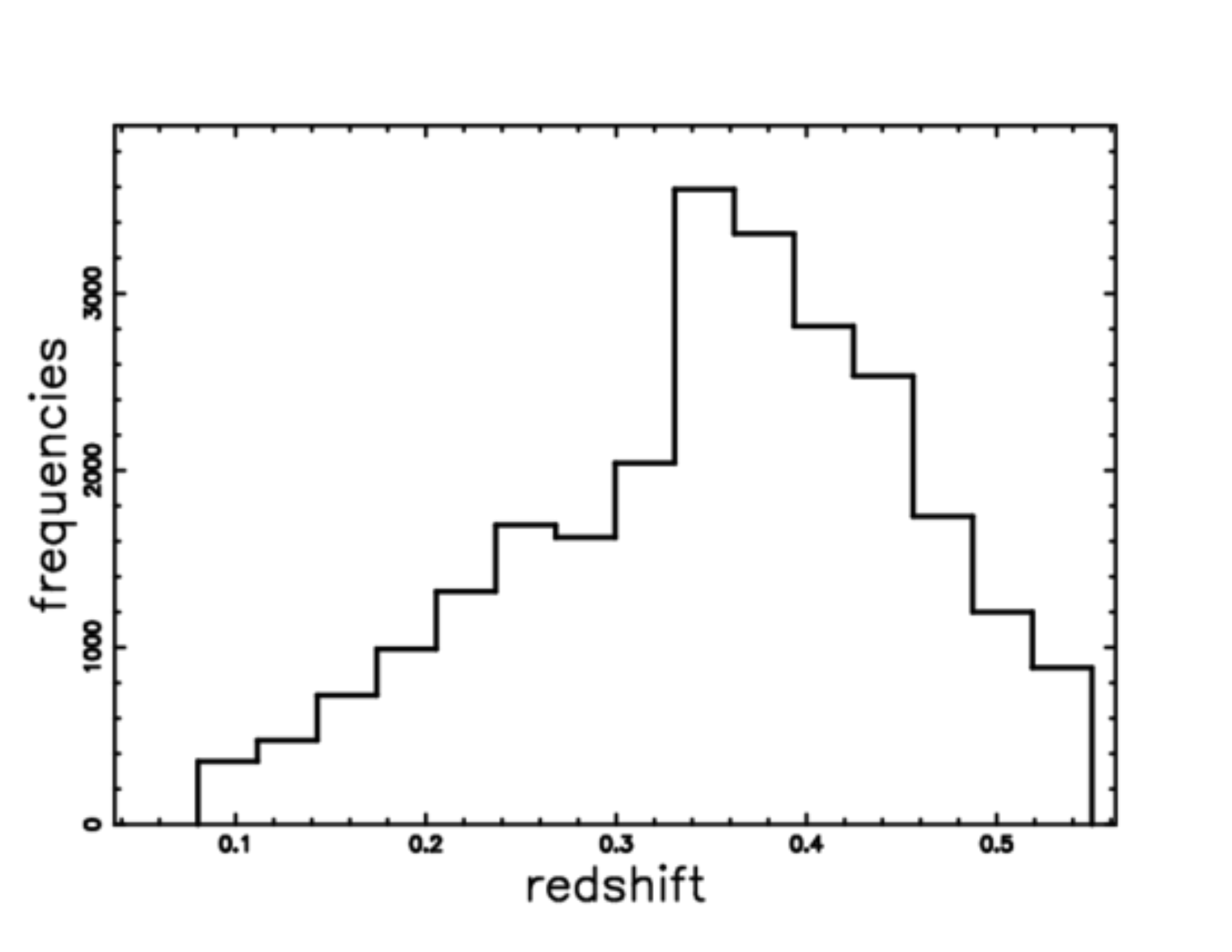}
\end{center}\caption
{
Histogram (step-diagram) of the number of clusters in redMaPPer
as a function of the redshift.
}
          \label{max_clusters_red}%
    \end{figure}

\begin{figure}\begin{center}
\includegraphics[width=7cm]{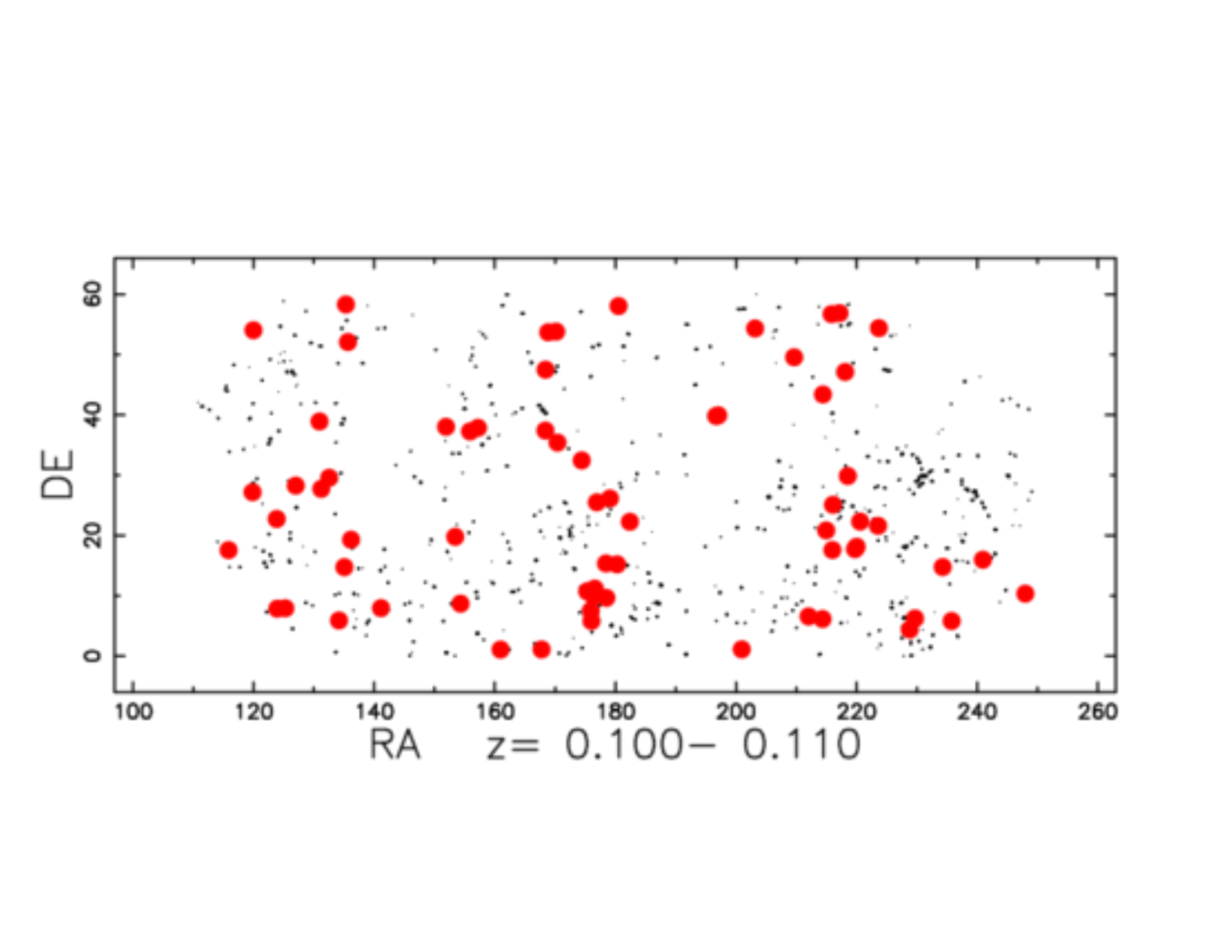}
\end{center}\caption
{
Cut  of the 3D spatial distribution  of
redMaPPer  catalog at $0.1 < z< 0.11$:
galaxies (black points) and clusters (red circles).
}
          \label{redmapper_clusters}%
    \end{figure}

\begin{figure}\begin{center}
\includegraphics[width=7cm]{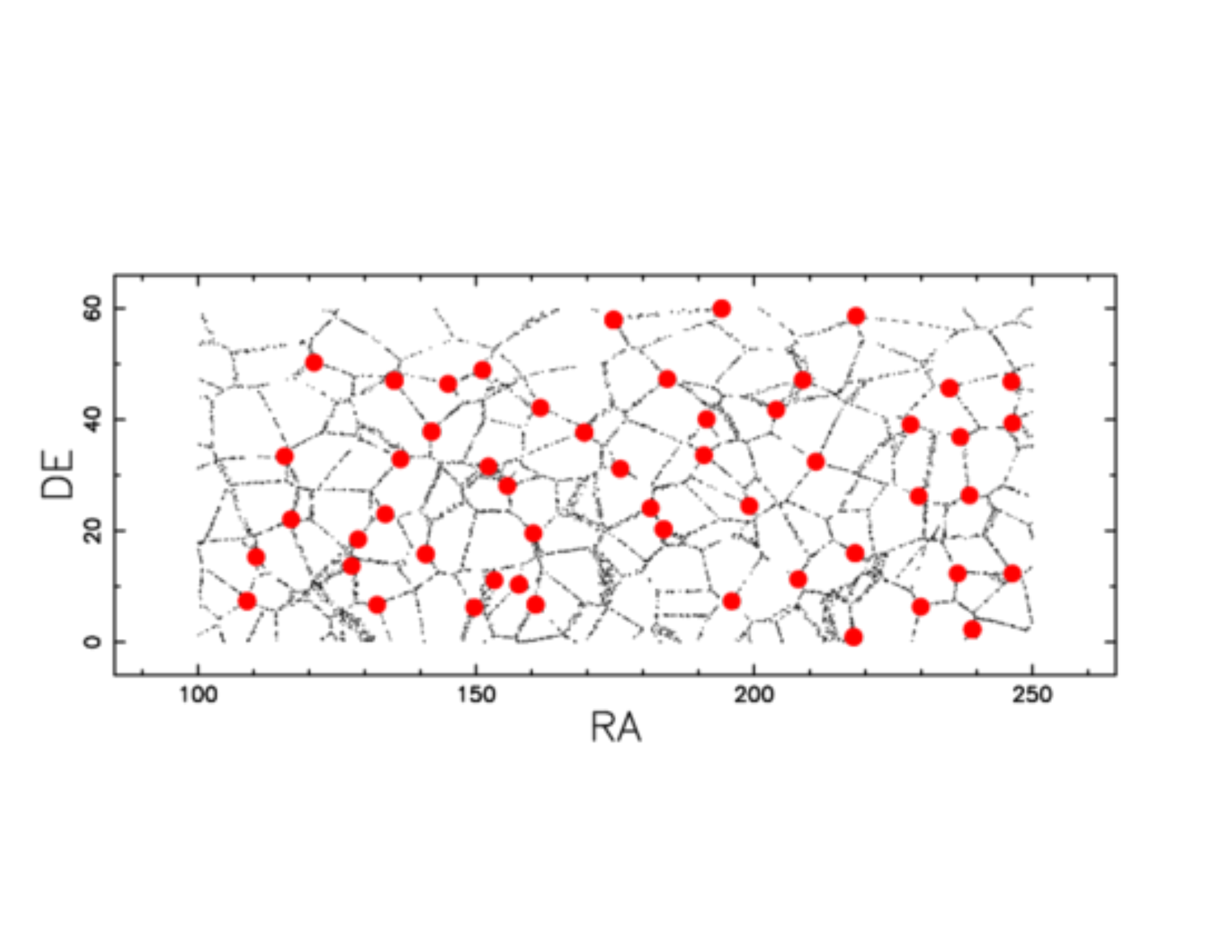}
\end{center}\caption
{
$V_{P,s}(2,3)$ (black points) and  spherical nodes (red circles)
at
$0.1 < z< 0.11$ or  $480\,Mpc < D_L 484\, < \,Mpc$.
The numbers  of galaxies and clusters are the same as in
Figure \ref{redmapper_clusters}.
}
          \label{redmapper_rectangular_due}%
    \end{figure}

The  number of clusters of the redMaPPer  catalog
as a function  of  the redshift  has a maximum
at $z \approx  0.346$, see
Figure \ref{max_clusters_red}.
Figure \ref{redmapper_clusters} reports the
galaxies and clusters of
the redMaPPer  catalog.
Figure \ref{redmapper_rectangular_due} reports
the same number of $V_{P,s}(2,3)$
and the spherical  nodes
with the same parameters as in
Table\ref{parameterspseudo}.


\section{Conclusions}

{\bf Adopted Cosmology}
The local universe can be considered Euclidean
up  to $z \approx 0.2$, see Figure \ref{distances}.
At higher values of redshift, we need a
a luminosity distance versus redshift  relationship in
$\Lambda$CDM cosmology, see equation (\ref{dlz}),
and a redshift versus luminosity distance  relationship,
see equation (\ref{zdl}).

{\bf Clusters of galaxies}
The geometrical nature for the clusters of galaxies
is introduced in the framework of the intersection
generated by a sphere with  the faces of the Voronoi diagrams.
At the same time, the exact number of galaxies and clusters
as function of the redshift is regulated
by  photometric rules;
i.e., a maximum in the  number of galaxies,
see  Figure  \ref{maximum_2mrs} for the 2MRS catalog.
This maximum is also visible for the
number of clusters as a function
of the redshift, see Figure \ref{max_clusters_red}
for  the redMaPPer  catalog.
A theoretical result for the spatial distribution of
galaxies  and
clusters  for    the redMaPPer  catalog
is reported in Figure \ref{redmapper_rectangular_due}.

\section*{Acknowledgments}

The 2MASS Redshift Survey is  available
at  \url{http://cdsweb.u-strasbg.fr/},
the VizieR catalogue access tool, CDS,
Strasbourg, France.


\end{document}